# Searches for Extra Dimensions at the Tevatron

Müge Karagöz Ünel

for the CDF and DØ Collaborations

*Northwestern University, 2145 Sheridan Rd., Illinois, 60208, USA*

**Abstract.** Models in which gravity and/or Standard Model gauge bosons propagate in more than three spatial dimensions have implications that can be tested at current colliders. In this paper, we report on the results from searches for extra dimensions at the two Tevatron experiments, CDF and DØ, which utilize up to 200 pb$^{-1}$ of p$\bar{\text{p}}$ collision data from Run II taken at $\sqrt{s}$ = 1.96 TeV between spring 2002 and fall 2003.

## INTRODUCTION

The hierarchy problem between the Planck scale ($M_{Pl}$) and the electroweak scale ($M_{EW}$) ($10^{16}$) has motivated a number of extensions to the Standard Model (SM) [1]. Models in which particles can propagate in additional spatial dimensions have been proposed [2]. In the model proposed by Arkani-Hamed, Dimopoulos and Dvali (ADD) [3], SM particles are confined on to a spatial 3-D membrane and gravity is allowed to propagate in the $n$ extra dimensions (ED). In the (4+$n$)-D world which corresponds to the fundamental Planck scale, $M_S$ (sometimes called as the string scale), gravity is as strong as other gauge forces, however, it is weak in the Planck scale ($M_{Pl} \sim 10^{19}$ GeV) in 4-D. The relation between the two Planck scales is governed through the size of the extra dimension, $R$ ($M_{Pl}^2 \sim R^n M_S^{n+2}$). Large extra dimensions (LED) is satisfied by the ADD model. Assuming $M_S$ is near the TeV range, R becomes as large as $\sim 10^{18}$ km, which is already ruled out by the classical description of gravitational force. The LED are compactified and the gravitational field in the (4+$n$)-D space can be represented in series of Kaluza-Klein (KK) towers.

Randall and Sundrum (RS) propose a non-factorizable geometry in 5-D space to addresses the hierarchy problem. The extra dimension is warped by an exponential factor, $\exp(-2kr_c\phi)$, where $k$ is the curvature scale and $r_c$ is the size of the 5th dimension [4] and the model predicts graviton resonances. The properties of the RS Kaluza-Klein states are usually expressed in terms of a dimensionless coupling parameter, $c = k/M_{Pl}$, the relative strength of $k$ to the effective Planck scale. A third model, TeV$^{-1}$ ED, is proposed to achieve gauge coupling unification at a scale much lower than the GUT scale [5, 6]. In this model, matter resides on a p-brane (p > 3), with chiral fermions confined to the ordinary 3-D world and SM gauge bosons also propagating in the extra dimensions, all of which are internal to the p-brane. The SM gauge bosons that propagate in the ED are equivalent to $n$ KK towers with masses $M_n = \sqrt{M_0^2 + n^2/R_C^2}$, where $R$ is the size of the compact dimension of the compactification scale, $M_C$, and $R = M_C^{-1}$.

**TABLE 1.** Summary of Tevatron searches discussed in the paper.

| Signature | | Model |
|---|---|---|
| Graviton emission | jets + $\not{E}_T$ | ADD |
| Graviton exchange | dilepton, diphoton | ADD, RS |
| Gauge boson exchange | dielectron | TeV$^{-1}$ ED |

The existence of extra dimensions can manifest itself in $p\bar{p}$ collisions through exchanges of KK towers of gravitons yielding difermion or diboson final states or through production of real gravitons along with a recoiled photon or jet yielding a high unbalanced transverse energy ($\not{E}_T$) due to the missing graviton. The CDF [7] and DØ [8] experiments of Tevatron Run II [9] have searched for ED models using up to about 200 pb$^{-1}$ of $p\bar{p}$ collisions at $\sqrt{s}$ = 1.96 TeV collected between spring 2002 and fall 2003. Table 1 outlines the signatures and the models probed at the Tevatron for the above-mentioned data-taking period.

## GRAVITON EXCHANGE SEARCHES

### ADD Virtual Graviton Exchange

In the ADD model, the separation between the KK towers of the graviton is small due to the large size of the extra dimensions. The SM diphoton production and the Drell-Yan pair production of leptons are modified in the existence of a KK exchange through the interference terms resulting in a continuum mass spectrum (Figure 1). The effect of KK exchange is expressed through an "effective" cross section parameterized by a single variable, $\eta_G = \mathscr{F}/M_S^4$, where $\mathscr{F}$ is a dimensionless parameter that reflects the dependence on the number of extra dimensions. The effective cross section is linear in $\eta_G$ for the interference term and quadratic in $\eta_G$ for the pure graviton exchange term. Different formalisms exist for interpreting the effects of LED through $\mathscr{F}$ [10, 11, 12]. These are listed in Eq. (1).

$$\mathscr{F} = \begin{cases} 1 & \text{(GRW)}, \\ \log(\frac{M_S^2}{M^2}), (n=2), \quad \frac{2}{n-2}, (n>2) & \text{(HLZ)}, \\ \frac{2\lambda}{\pi} = \pm\frac{2}{\pi} & \text{(Hewett)}. \end{cases} \quad (1)$$

Both CDF and DØ experiments have searched for the large ED signal in the virtual graviton exchange mode. The DØ experiment's most sensitive channel is "diEM" sample, which is a combination of dielectron and diphoton channels to maximize detection effi-ciency [13]. The basic selection requirements are two electromagnetic (EM) objects in the calorimeter ($E_T > 25$ GeV) with track isolation. One EM object is central ($|\eta| < 1.1$), the other can be central or in the end-cap calorimeter ($1.5 < |\eta| < 2.4$). The main background contributions to the analysis are SM Drell-Yan and diphoton production and one or more jets misidentified as an EM object (instrumental background). The systematic uncertainty for the background is 7–20%, dominated by the statistical and systematic uncertainties of the instrumental background. The signal systematic uncertainty is 12%, which includes errors on signal acceptance and efficiency, choice of parton distribution functions (PDF) and the higher order correction factor, $K_f$ (=1.3). The uncertainty is dominated by the 10% error on the $K_f$. The DØ diEM invariant mass spectrum for the candidate events in data and its comparison with the predicted background is shown in Figure 1. Similar comparison for the diEM $\cos\theta^*$ distribution, where $\theta^*$ is the scattering angle in the center-of-mass frame, is also shown in the same figure. The data agrees with the background expectation.

**TABLE 2.** 95% C.L. lower limits on $M_S$ in TeV for LED with the 200 pb$^{-1}$ DØ Run II diEM sample and the combined limits for Run I and II.

|  | GRW | HLZ | | | | | | Hewett |
|---|---|---|---|---|---|---|---|---|
|  |  | $n=2$ | $n=3$ | $n=4$ | $n=5$ | $n=6$ | $n=7$ | $\lambda=+1$ |
| Run II | **1.36** | 1.56 | 1.61 | 1.36 | 1.23 | 1.14 | 1.08 | 1.22 |
| Run I+II, combined | **1.43** | 1.67 | 1.70 | 1.43 | 1.29 | 1.20 | 1.14 | 1.28 |

DØ employs a search strategy which involves a 2-D fit to the invariant mass and (unsigned) $\cos\theta^*$ spectra. Using a Bayesian likelihood fitting technique, 95% Confidence Level (C.L.) lower limits on the effective cross section parameter, $\eta_G$, are placed and translated onto lower limits on $M_S$. The results of the DØ searches using diEM channel are listed in Table 2. A combination with the Run I [14] analysis result gives lower bounds on $M_S$ as high as 1.43 TeV (GRW).

DØ has also results from the dielectron and dimuon channels, where both channels use the same limit setting technique as in the diEM channel. The dielectron channel selection requirements are similar to the diEM analysis with the exception of a modification to allow for at least one EM cluster to have a matching track and the tracking isolation cut is removed to increase efficiency [15]. The lower limit obtained using dielectron channel is 1.11 TeV in the GRW formalism. The LED virtual effects in the dimuon channel is searched using 100 pb$^{-1}$ data [16]. An upper limit of 0.88 TeV is placed on $M_S$ in the GRW formalism. The dilepton LED results from the DØ experiment is outlined in Table 4.

The CDF experiment has also searched for a LED signal in the dielectron channel using 200 pb$^{-1}$ of data [17]. The requirements in the analysis are two electrons ($E_T > 25$ GeV), isolated in the calorimeter, with at least one electron matched to a track and present in the central calorimeter ($|\eta| < 1.0$). The second electron can also be in the plug calorimeter ($1.0 < |\eta| < 3.0$). The main backgrounds to CDF dielectron signal are from the Drell-Yan process and from misidentified electrons in QCD multijet events. Other SM processes like diboson and $t\bar{t}$ production are also taken into account, although their contribution for the high mass search region is negligible. The comparison of the observed dielectron invariant mass spectrum to the predicted background is shown in Figure 2. The main sources of systematic uncertainties are choice of PDF, energy scale and resolution for the acceptance and efficiency, luminosity and background shape and normalization errors. The overall signal uncertainty is about 8%. The background due to misidentified jets is about 50%. In the absence of an excess in the observed data, CDF proceeds to set limits on $M_S$, using a Bayesian binned likelihood method on the dielectron invariant mass spectrum. The 95% C.L. limits on $\eta_G$, where the convention is $\eta_G = \lambda_{Hewett}/M_S^4$, are used to constrain the string scale. Table 3 summarizes CDF dielectron limit results for $M_S$.

**TABLE 3.** 95% C.L. lower limits on $M_S$ in TeV for LED model with 200 pb$^{-1}$ CDF dielectron sample.

| GRW | HLZ | | | | | Hewett | |
|---|---|---|---|---|---|---|---|
| | $n=3$ | $n=4$ | $n=5$ | $n=6$ | $n=7$ | $\lambda=-1$ | $\lambda=+1$ |
| **1.11** | 1.17 | 0.99 | 0.89 | 0.83 | 0.79 | 0.99 | 0.96 |

## Gauge Boson Exchange in TeV$^{-1}$ ED

The cross section for Drell-Yan pair production including TeV$^{-1}$ ED effects from the higher order KK-states of the SM gauge bosons are parametrized in a form similar to that of LED effective cross section, where the parameter is $\eta_C = \pi^2/3M_C^2$ and $M_C$ is the compactification scale of the TeV$^{-1}$ ED. DØ has searched for TeV$^{-1}$ ED signatures for the case of a single ED, using 200 pb$^{-1}$ data in the dielectron channel [15]. The analysis uses the same dielectron sample as the LED search described earlier. Figure 2 shows the dielectron candidates in the DØ data and the predicted background. Also shown in this figure is the effect of the TeV$^{-1}$ ED signal for $\eta_C = 5.0$ TeV$^{-2}$. Systematic uncertainty on the signal is taken similar to the LED search, and is 12%. A 10% systematic uncertainty is estimated for background processes, dominated by the uncertainties in the instrumental background contribution. Limits on $M_C$ are set using the 2-D fitting apparatus developed for LED search, via a replacement of $\eta_G$ with $\eta_C$. The lower limit on the compactification scale is obtained as $M_C > 1.12$ TeV. The combined indirect search limit from LEP is $M_C > 6.6$ TeV and the global limit is $M_C > 6.8$ TeV at 95% C.L. [18].

## RS Graviton Resonance Searches

The CDF experiment has searched for a RS graviton signal of the first excited state in both dilepton and diphoton channels[1]. The dielectron sample used for the search is the same as the CDF LED search. The CDF search for RS graviton in the dimuon channel is performed using 200 pb$^{-1}$ data, similar to that of the dielectron channel. The muon pair selection is based on requiring two muons isolated in the calorimeter ($p_T > 20$ GeV/c). One muon should be central in the CDF detector ($|\eta| < 1.0$), the other can be present in the tracking region, $|\eta| < 1.5$. The backgrounds in the dimuon channel are the irreducible Drell-Yan, misidentified jets and cosmic rays. Various other SM processes are also taken into account as contributors to the total background, as seen in Figure 3. The figure shows the comparison of the observed data to the background estimates for the dimuon mass spectrum. The uncertainty in the signal is about 8% and the background uncertainties are estimated to be 5% (direct dimuon final states from SM processes) and 20–30% (fakes and cosmic rays).

---

[1] As of writing this paper, DØ presented results on RS graviton searches in the diEM channel using 200 pb$^{-1}$ data [19].

**TABLE 4.** 95% C.L. lower limits on $M_S$ in TeV for LED with DØ dielectron and dimuon sample.

|  | GRW |
|---|---|
| $ee$ (200 pb$^{-1}$) | **1.11** |
| $\mu\mu$ (100 pb$^{-1}$) | **0.88** |

CDF recently increased its sensitivity to RS gravitons with the addition of the diphoton channel [20]. The result presented utilizes also 200 pb$^{-1}$ data and the photon pair candidates ($E_T > 15$ GeV, $M_{\gamma\gamma} > 30$ GeV/c$^2$) are constrained to be central ($|\eta| < 1.0$) and isolated. The main background contributions are from SM hard-scattered diphoton production and from events in which one or more jets can be misidentified as a photon. The total uncertainty for the signal is about 15% and is 20–25% for the background depending on the diphoton mass. The dominant uncertainty in this analysis is from the dependence of the jet background on the selection requirements. The observed diphoton spectrum is compared to the prediction in Figure 3.

The CDF dilepton/diphoton data is in agreement with the Standard Model predictions. Therefore, limits are placed on a RS graviton particle of the first excited state. The CDF dilepton limits are placed by a Bayesian binned likelihood method using the dilepton invariant mass spectrum. The diphoton limits are placed via a Bayesian technique as a function of RS graviton reference masses and using a $3\sigma$ mass window around each reference mass. Table 5 summarizes the CDF RS graviton search results. The $\sigma \cdot BR$ limits are interpreted as limits on a spin-2 resonance particle, which are further used to constrain the RS graviton mass as a function of k/M$_{Pl}$. The acceptance times efficiency of the diphoton channel is about half of that of the individual dilepton channels (the total dielectron efficiency to detect a 600 GeV/c$^2$ RS graviton is about 45%). Therefore, the $\sigma \cdot BR$ limits from the diphoton channel is not as sensitive for the same luminosity [20]. The stringent constraints to the RS graviton parameter space is from the diphoton channel to which the graviton has twice the *BR* as compared with its individual dilepton decays.

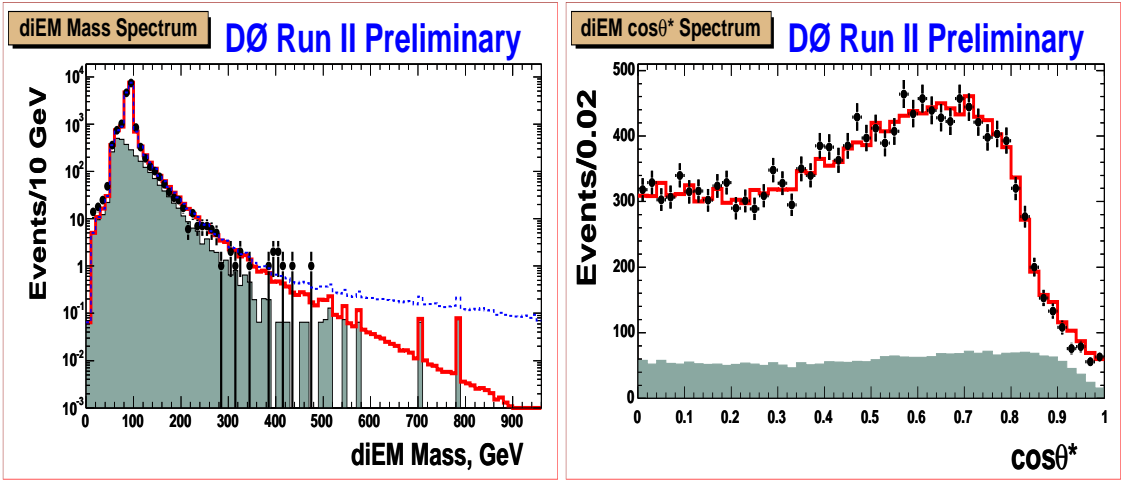

**FIGURE 1.** DØ diEM mass and $\cos\theta^*$ spectra for the candidates in data (points with error bars), for the misidentified jet background (filled histogram) and for the total expected background including the SM predictions (solid line). In the mass spectrum plot, the dashed line represents the effect of the LED signal for $\eta_G = 0.6$ TeV$^{-4}$.

## GRAVITON EMISSION SEARCHES

DØ has analyzed 85 pb$^{-1}$ of Run II data to search for evidence of a signal of LED with graviton emission signature in jet(s) + $\not{E}_T$ channel [21]. The analysis requires $\not{E}_T > 150$ GeV, a high $p_T$ leading jet in the central region ($p_T > 150$ GeV/c), no second jet with $p_T > 50$ GeV/c, no high $p_T$ leptons in the event, and an angular separation between the leading jet and $\not{E}_T$ ($\Delta\Phi$). The main SM backgrounds come from Z+jets production (through invisible decays of Z) and a smaller contribution from W+jets production. The dominating background at low $\not{E}_T$ is misidentified QCD events. Figure 5 shows the total $\not{E}_T$ distribution before the second jet and $\Delta\Phi$ requirements are applied in the analysis data. The expected number of events is 100 ± 6 (stat.) ± 8 (theory). The current result is limited by the large Monte Carlo simulation and data jet energy scale uncertainties, which yields uncertainties of 20% for the signal efficiency and +50%, -30% for the background prediction. Total signal efficiency is 5% and 63 events in data are observed. In the absence of excess in data with respect to the expectations from background processes, lower limits on $M_D$ are placed using the LEP CLs approach [22]. The corresponding upper limits on the number of signal events is 84 event for an expected limit of 111.4 (median) or 123.8±28 events (average) from ensemble tests. Figure 6 shows the lower limits on $M_D$ as a function of $n_D$, together with the limits from Run I CDF [23] ($K_f = 1.0$) and Run I DØ [24] and the current most stringent limits from LEP [25]. Once the uncertainties settled, the analysis will be more competitive.

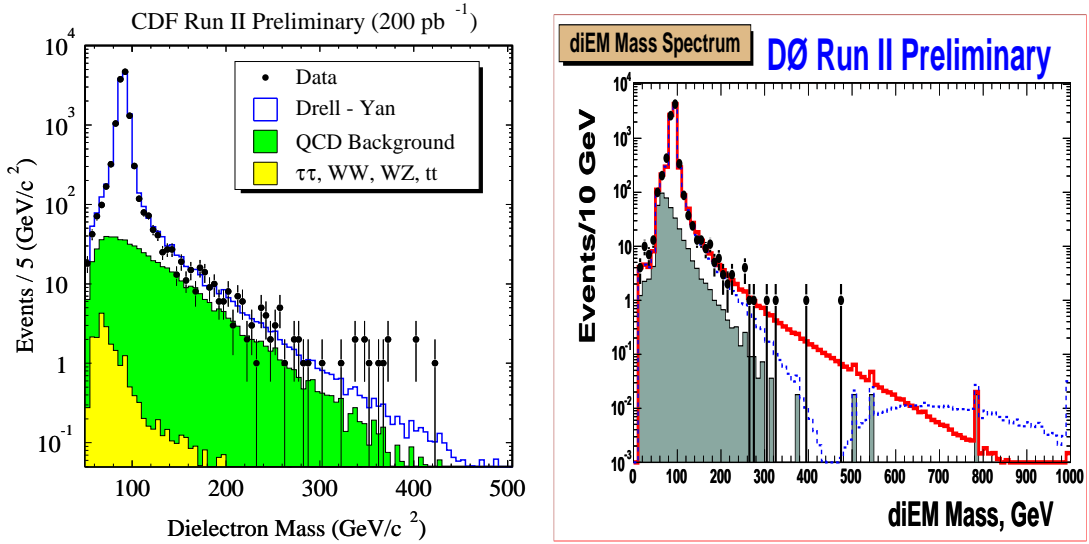

**FIGURE 2.** The CDF (left) and DØ (right) dielectron invariant mass spectra. The CDF data (points) are compared with the total background predictions (solid line) from Drell-Yan (unfilled histogram), multijet QCD background (dark) and various SM processes as listed (light). The DØ data (points) are compared with the total predicted background (solid line) as the sum of the instrumental background (filled histogram) and the Drell-Yan background. The plot on the right also represents TeV$^{-1}$ ED signal contribution for $\eta_C = 5.0$ TeV$^{-2}$ added to the predicted background (dashed line).

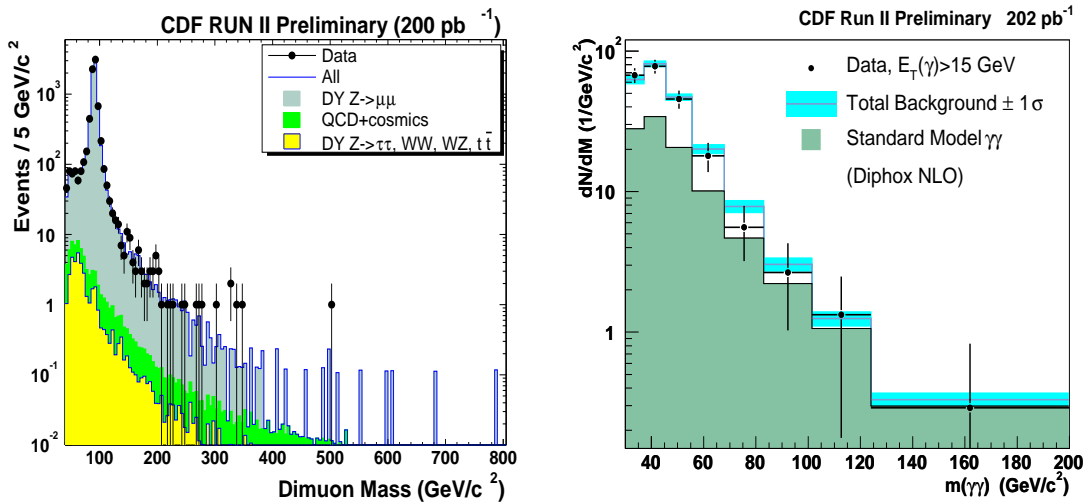

**FIGURE 3.** Observed CDF dimuon (left) and diphoton (right) mass spectra compared to the predicted backgrounds as listed in the text. The diphoton spectrum has three events above $M_{\gamma\gamma} = 200$ GeV/c$^2$; the highest one being at 405 GeV/c$^2$.

## SUMMARY AND CONCLUSION

With the Tevatron and its detectors performing very well, the CDF and DØ experiments have analyzed large samples of p$\bar{\text{p}}$ data taken since spring 2002. The experiments are actively engaged in extra dimension searches in dielectron, diphoton and dimuon channels as well as jets + $\not{E}_T$ channel. No evidence of higher dimensional models have been found yet. Preliminary 95% C.L. limits have been placed on the parameter space of such models. Current LED limits of the Tevatron experiments exceed published limits of previous direct searches. The collaborations also pioneer the search strategies and exploration of some models. Additional extra dimension signatures are being explored, such as $\gamma + \not{E}_T$. The sensitivity to ED searches will increase with more Run II data. Near the end of the decade, experiments at LHC will explore extra dimension models up to multi-TeV scales [26]. In case of no discovery, parameter spaces for ED models will be greatly restrained and some models will possibly be completely ruled out.

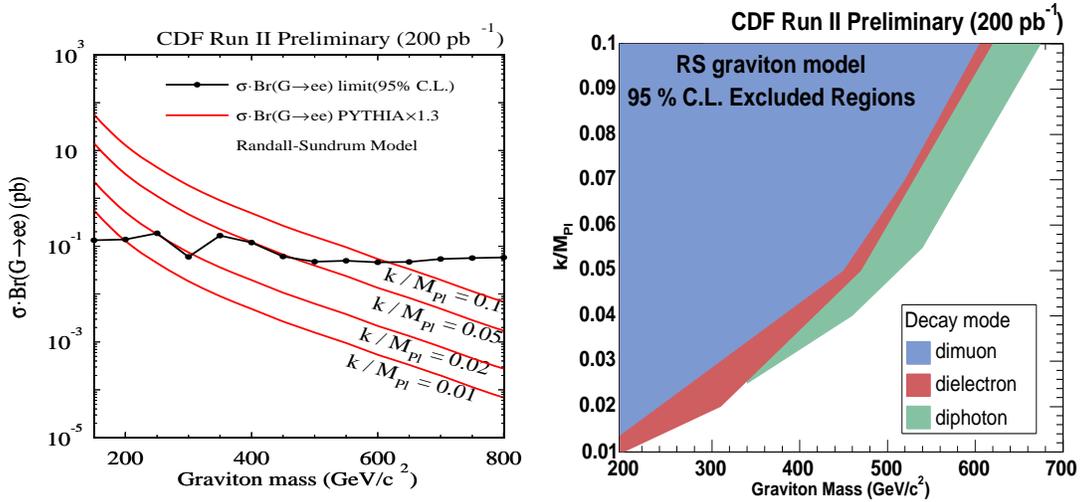

**FIGURE 4.** Upper $\sigma \cdot BR$ limits on RS graviton production in dielectron channel (left) and excluded parameter space in dilepton and diphoton channels at 95% C.L. (right).

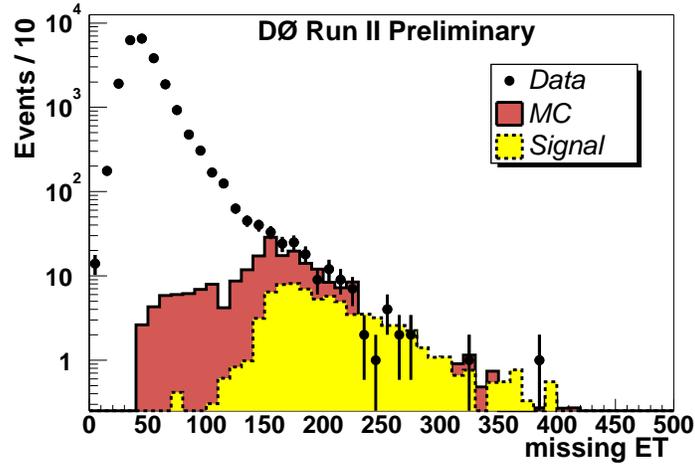

**FIGURE 5.** $\displaystyle{\not}E_T$ distribution for graviton emission searches, for data (points with error bars), non-QCD SM background (full histogram) and a signal simulation ($n_D = 6$, $M_D = 0.7$ TeV; dashed histogram). QCD background dominates at low values of the spectrum.

**TABLE 5.** 95% C.L. upper limits on $\sigma \cdot BR(G \to \ell^+\ell^-, \gamma\gamma)$ and lower limits on $M_G$ with 200 pb$^{-1}$ CDF dilepton and diphoton sample for k/$M_{Pl}$ = 0.1.

|  | $ee$ | $\mu\mu$ | $\gamma\gamma$ |
|---|---|---|---|
| $\sigma \cdot BR$ (pb) ($M_G > 600$ GeV/c$^2$) | $\sim 50$ | $\sim 50$ | $\sim 100$ |
| $M_G$ (GeV/c$^2$) | 620 | 605 | 675 |

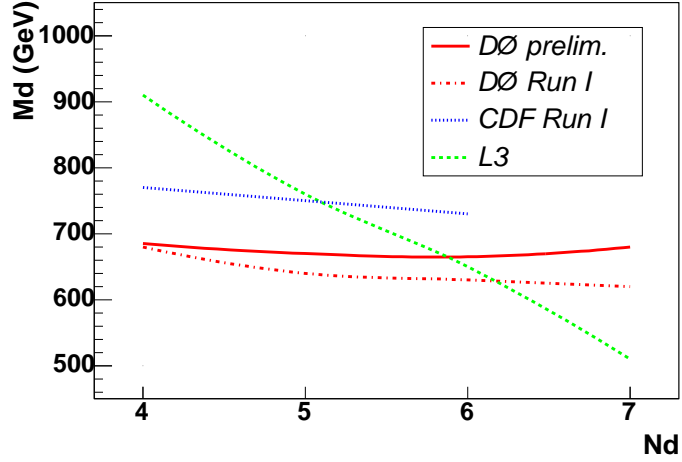

**FIGURE 6.** Lower limits on $M_D$ for various number of extra dimensions $n_D$ for DØ Run II (solid line), DØ Run I (dashed-dotted), CDF Run I (dotted) and LEP (dashed).

## ACKNOWLEDGMENTS

The author is grateful to Stephan Lammel for useful discussions.

## REFERENCES

1. Dawson, S., Perspectives of Standard Model, Proceedings, this conference (2004).
2. Burdman, G., Theories with Extra Dimensions, Proceedings, this conference (2004).
3. Arkani-Hamed, N., Dimopoulos, S., and Dvali, G., *Phys. Lett.*, **B429**, 263 (1998).
4. Randall, L., and Sundrum, R., *Phys. Rev. Lett.*, **83**, 3370–3373 (1999).
5. Dienes, K., Dudas, E., and Gherghetta, T., *Nucl. Phys.*, **B537**, 47 (1999).
6. Pomarol, A., and Quiros, M., *Phys. Lett.*, **B438**, 255 (1998).
7. Blair, R. et al., the CDF Collaboration, The CDF-II Detector: Technical Design Report, Fermilab-Pub-96-390-E, 1996, 234pp; Abe. F. et al., the CDF Collaboration, *Nucl. Instrum. Methods*, **A271**, 387 (1988).

8. Tuts, P. M. et al., the DØ Collaboration, *Nucl. Phys. Proc. Suppl.*, **32**, 29 (1993); Abachi, S. et al., the DØ Collaboration, *Nucl. Instrum. Methods*, **A338**, 185 (1994).
9. Lipton, R., Tevatron Detector Upgrades, Proceedings, this conference (2004).
10. Guidice, G. F., Rattazzi, R., and Wells, J. D., *Nucl. Phys.*, **B544**, 3–38 (1999), and revised version hep-ph/9811291, URL http://arxiv.org/abs/hep-ph/9811291.
11. Han, T., Lykken, J., and Zhang, R., *Phys. Rev.*, **D59**, 105006 (1999).
12. Hewett, J. L., *Phys. Rev. Lett.*, **82**, 4765–4768 (1999).
13. The DØ Collaboration, Search for Large Extra Dimensions in the Dielectron and Diphoton Channels with 200 pb$^{-1}$ of Run II Data, DØ Public Note 4336 (2004), URL http://www-d0.fnal.gov/Run2Physics/WWW/results/NP/N01/N01.pdf.
14. Abbot, B. et al., the DØ Collaboration, *Phys. Rev. Lett.*, **86**, 1156 (2001).
15. The DØ Collaboration, Search for Large and TeV$^{-1}$ Extra Dimensions in the Dielectron Channel with 200 pb$^{-1}$ of Data, DØ Public Note 4349 (2004), URL http://www-d0.fnal.gov/Run2Physics/WWW/results/NP/N02/N02.pdf.
16. Vilar, R., Non SUSY Searches at the Tevatron, Proceedings, The XXXIXth Rencontres de Moriond, La Thuile, Italy, March 28 – April 4, 2004; an update of this analysis with 250 pb$^{-1}$ is available as DØ Public Note 4576 (2004), URL http://www-d0.fnal.gov/Run2Physics/WWW/results/NP/N19/N19.pdf.
17. Karagoz Unel, M., Searches for New Physics in High Mass Dileptons at CDFII, talk given at Phenomenology Symposium, University of Wisconsin-Madison, April 26–28, 2004, URL http://www.pheno.info/symposia/pheno04; also available in [16].
18. Cheung, K., Landsberg, G., *Phys. Rev.*, **D65**, 076003 (2002).
19. The DØ Collaboration, Search for Randall-Sundrum Gravitons in the Dielectron and Diphoton Final States with 200 pb$^{-1}$ of Run II Data with the DØ Detector, DØ Public Note 4533 (2004), URL http://www-d0.fnal.gov/Run2Physics/WWW/results/NP/N17/N17.pdf.
20. The CDF Collaboration, Search for a High-Mass Diphoton State and Limits on Randall-Sundrum Gravitons at CDF, CDF Public Note 7098 (2004), URL http://www-cdf.fnal.gov/physics/exotic/public-notes/cdf7098_RSG_dipho_Search202.pdf; an update of this analysis with 345 pb$^{-1}$ is available as a revision of CDF Public Note 7098 (2004).
21. The DØ Collaboration, Search for Large Extra Spatial Dimensions in Jets + $\not{E}_T$ Topologies, DØ Public Note 4400 (2004), URL http://www-d0.fnal.gov/Run2Physics/WWW/results/NP/N06/N06.pdf.
22. Read, A. L, Modified Frequentist Analysis of Search Results (the CLs Method), in *CERN Report 2000-005*, edited by F. James, L. Lyons and Y. Perrin, 2000, p. 81.
23. Acosta, D. et al., the CDF Collaboration, *Phys. Rev. Lett.*, **92**, 121802 (2004).
24. Abazov, V. M. et al., the DØ Collaboration, *Phys. Rev. Lett.*, **90**, 251802 (2003).
25. Achard, P. et al., the L3 Collaboration, *Phys. Lett.*, **B587**, 16–32 (2004).
26. Tovey, D., New Physics Prospects at the LHC, Proceedings, this conference (2004).